\documentclass[journal]{IEEEtran}

\ifCLASSINFOpdf
\else
   \usepackage[dvips]{graphicx}
\fi
\usepackage{url}
\usepackage{romannum}

\usepackage{graphicx}
\usepackage{amsmath,graphicx,mathtools,nth,hhline,multirow,color,cite}
\usepackage{arydshln}

\usepackage{booktabs}
\usepackage{cite}
\usepackage{siunitx}

\usepackage{xcolor}

\usepackage[symbol]{footmisc}

\begin{document}

\title{Drum-Aware Ensemble Architecture for Improved Joint Musical Beat and Downbeat Tracking}

\author{Ching-Yu Chiu, \IEEEmembership{Student Member, IEEE}, Alvin Wen-Yu Su, and Yi-Hsuan Yang, \IEEEmembership{Senior Member, IEEE}
\thanks{Ching-Yu Chiu is with the Graduate Program of Multimedia Systems and Intelligent Computing, National Cheng Kung University and Academia Sinica, Taiwan (e-mail: sunnycyc@citi.sinica.edu.tw).}
\thanks{Alvin Wen-Yu Su, is with the Department of Computer Science and Information Engineering, National Cheng Kung University, Taiwan (e-mail: alvinsu@mail.ncku.edu.tw).}
\thanks{Yi-Hsuan Yang, is with Yating Music Team, Taiwan AI Labs, Taiwan. He is also with Research Center for IT Innovation, Academia Sinica, Taiwan (e-mail: yang@citi.sinica.edu.tw).}}

\maketitle

\begin{abstract}
This paper presents a novel system architecture that integrates blind source separation with joint beat and downbeat tracking in musical audio signals.  The source separation module segregates the percussive and non-percussive components of the input signal, over which beat and downbeat tracking are performed separately and then the results are aggregated with a learnable fusion mechanism.
This way, the system can adaptively determine how much the tracking result for an input signal should depend on the input's percussive or non-percussive components.
Evaluation on four testing sets that feature different levels of presence of drum sounds shows that the new architecture consistently outperforms the widely-adopted baseline architecture that does not employ source separation.

\end{abstract}


\IEEEpeerreviewmaketitle

\section{Introduction}

\IEEEPARstart{B}{eats} and downbeats, usually referred to as the sequence of times to tap to when listening to a piece of music, are fundamental information for analyzing and understanding music \cite{ballroom2,Bock2014b,davies11spl,Holzapfel2012b,durand17taslp}. 
Besides being important in its own right, automatic beat and downbeat tracking holds downstream applications in tasks such as music transcription \cite{benetos19spm}, structural segmentation,  
automatic accompaniment \cite{burloiu16smc}, 
and music generation \cite{huang20mm}. 
As such, beat/downbeat tracking has been an important topic in signal processing and music information retrieval. 
And, along with the development of deep learning based techniques, the performance of beat/downbeat tracking has been improved greatly over the recent years \cite{Bock2016d,Bock2016,Fuentes2018,Bock,giorgi20ismir,cano20ismir,pedersoli20ismir}.


\begin{figure}
\centerline{\includegraphics[width=\columnwidth]{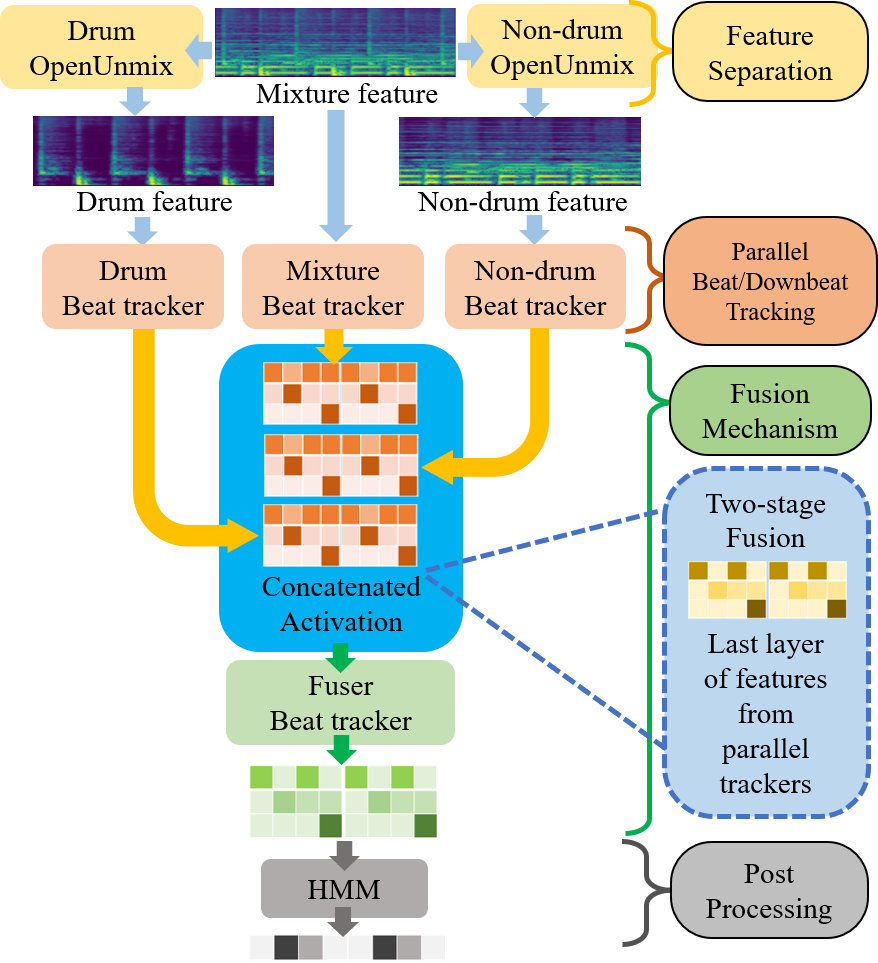}}
\caption{Overview of the proposed drum-aware ensemble architecture.}
\label{fig:sys_overview}
\end{figure}

Research on blind monaural source separation, which concerns with segregating the sound sources 
involved in a monaural audio recording, has also seen remarkable progress in recent years thanks to deep learning \cite{Stoter2019,Hennequin2019,liu19ijcai,chiu20mmsp}.  
Although not yet widely noted in recent work, we conjecture that source separation is mature enough to be integrated to a beat/downbeat tracking model, with the parameters of  both  optimized jointly, to improve the performance of the latter.
We are in particular interested in the case of using a source separation module that separates an input audio to merely two streams---a ``drum'' component and a ``non-drum'' component---and then learning to track beats and downbeats also over the resulting two streams, which are supposed to have the same metrical structure as the original input.
The rationale behind is that, with or without the drum sounds, human can steadily tap to a song by switching their attention between the percussive and non-percussive parts \cite{Gouyon_acoustic_cue}.  
Presenting not only the original input but also the separated components to the machine allows it to develop specialized trackers 
for three types of inputs (i.e., original mixture, drum and non-drum), whose predictions can be fused later on adaptively depending on the acoustic characteristics of the input (see Fig. \ref{fig:sys_overview} for an illustration and Section \ref{sec:methodology} for details).
Doing so endows the machine with the ability to shift its attention to non-conventional parts of the music, and may in turn improve its tracking accuracy across signals that feature different usages of the drum sounds.

It is a recurring observation (e.g., \cite{Fuentes2018}) that the performance of beat/downbeat tracking can vary a lot across different test sets. 
We conjecture that this is partly because
the drum sounds can play fairly different roles in different songs. 
For example, while acting as a timekeeper in many cases (e.g., in Pop and Rock), drum sounds can be more creative in Jazz and Funk, while be absent in classical or choral music \cite{Jia2019}.  
Using a single network to fit it all, as done in many recent work (e.g., \cite{Bock2016d,Bock2019}),
may lead to a model that relies on features that 
do not necessarily present in all types of  audio signals. 
While there are many other factors that can limit the performance of beat/downbeat tracking (e.g., the presence of local tempo changes \cite{Bock,Grosche2010a}), the impact of the versatile role of the drum sounds has not been much studied in the literature, to our best knowledge. It is therefore our goal to quantify 
this factor.  
Accordingly, the evaluation of the proposed model architecture is carried out using four test sets that 
are large enough, and that 
feature different levels of presence of drum sounds, including a new dataset of classical piano music proposed lately \cite{asap-dataset}, as listed in Table \ref{tab:dataset_overview}.
Three of the four test sets are kept completely unseen to our training scheme.


We implement our source separation and beat/downbeat tracking modules based on the bidirectional long-short term memory (BLSTM) architectures of the respective SOTA, the Open-Unmix model \cite{Stoter2019} and the RNNDownBeatProcessor of the Madmom library \cite{Bock2016d,Bock2016}. 
We open source our code at \url{https://github.com/SunnyCYC/drum-aware4beat}.

\section{Related Work}
Early work on beat/downbeat tracking tends to target on a one specific genre or style 
at a time with in-depth analysis of the signal properties \cite{Grosche2010a,Holzapfel2012b,Zapata2013a}. 
For example, the 
work from Goto and Muraoka \cite{Goto97musicunderstanding,GOTO1999311,Goto2001} presented a series of signal analysis methods for beat tracking in music with or without  drum sounds.
Harmonic-percussive source separation methods were also adopted to enhance the performance of dynamic programming-based beat tracking models \cite{McFee2014_onsetAGG,  percussivebeat2013}.

In more recent years, researchers began to focus on building systems capable of dealing with different kinds of music, via either feature or model design \cite{Durand2014}. 
The multi-model approach used to be mainstream. 
For example, both Durand \emph{et al.} \cite{Durand2015a,Durand2017a} and Krebs \emph{et al.} \cite{Krebs2016} tried to improve the performance and robustness of downbeat tracking with multiple complementary musical features that are fed to independent neural network modules, 
using the average of these modules
to yield the activation functions for the 
final postprocessing stage. 
For beat tracking, B{\"{o}}ck \emph{et al.} \cite{Bock2014b} also proposed a system that combines multiple recurrent neural networks specialised on certain musical styles with a model switcher. 

While beat and downbeat tracking used to be treated as separate tasks, multi-task models that jointly estimate beat, downbeat, and even also tempo, have become popular due to the  work of B{\"{o}}ck \emph{et al.} \cite{Bock2016d,Bock2019,Bock}.  
Joint beat/downbeat tracking models proposed since then tend to be a single model that aims to fit all the musical genres. 
Under the single-model framework, researchers employed data augmentation to deal with unseen or unpopular type of data \cite{Bock,Schreiber2018}.

The proposed model can be viewed as a multi-task multi-model system that uses learnable drum separation to divide the joint estimation of beats and downbeats in a  signal into  sub-problems, conquer them in parallel, and then fuse the result. 
This divide-and-conquer idea, which has not been explored before in  deep learning based beat/downbeat tracking to our best knowledge, can be easily integrated with any existing models, such as the BLSTM-based model \cite{Bock2016d} adopted here.


\begin{table}
\caption{The adopted datasets. Those on the top are split into 80\%, 10\%, 10\% for training, validation, and testing, we refer to the union of the test splits as ``Merged.'' The bottom three datasets are kept totally unseen as test sets. 
The rate on the last column shows the percentage of fragments with percussive sounds per dataset (see Section \ref{sec:exp:data} for details).}
\centering
\begin{tabular}{l|rrr}
\toprule
\multirow{2}{*}{Dataset}         & \multirow{2}{*}{\# song} & Total & Drum \\
  &  & duration & presence  \\
\midrule
RWC Classical  \cite{rwcdatabase}  & 54      & 5h 19m   &  0\%\\
RWC Jazz \cite{rwcdatabase}       & 50      & 3h 42m  &   29.8\%\\
RWC Music Genre  \cite{rwc_musicgenre} & 100     & 7h 20m  &   39.3\%\\
Ballroom \cite{ballroom1,ballroom2} & 685     & 5h 57m   &   59.1\%\\
Hainsworth \cite{hainsworth}      & 222     & 3h 19m   &  65.0\%\\
GTZAN \cite{gtzan1,gtzan2}          & 999     & 8h 20m  &   71.2\%\\
Carnatic  \cite{carnatic}    & 176     & 16h 38m   & 72.8\%\\
Beatles \cite{beatles}         & 180     & 8h 09m    & 75.8\%\\
RWC Popular  \cite{rwcdatabase}    & 100     & 6h 47m &    89.5\%\\
Robbie Williams \cite{robbiewilliams} & 65      & 4h 31m &    93.5\%\\
\midrule
Merged (10\% of the above datasets) & 263 & 6h 30m & 72.7\%\\
\midrule
ASAP  \cite{asap-dataset}           & 520     & 48h 07m &     0\%\\
Rock  \cite{rockdataset}           & 200     & 12h 53m  &  83.2\%\\
HJDB  \cite{Hockman2012}            & 235     & 3h 19m  &  99.2\%\\
\bottomrule
\end{tabular}
\label{tab:dataset_overview}
\end{table}

\section{Proposed Methods}
\label{sec:methodology}


Figure \ref{fig:sys_overview} depicts the proposed model architecture for joint beat and downbeat tracking. As shown on the right hand side of the figure, the model comprises four modules: feature separation, beat/downbeat tracking, fusion, and post processing. 
From the result of \emph{feature separation}, which performs source separation in a feature domain, the model uses an ensemble of three \emph{beat trackers} to get the activation likelihoods of beats, downbeats, and non-beats for the input audio mixture, the drum, and non-drum parts respectively. The \emph{fuser} aggregates the result from the three trackers (optionally with a two-stage fusion method; see later), and then a \emph{postprocessing} module, implemented as a hidden Markov model (HMM) \cite{Bock2016}, makes the final beat and downbeat prediction. 

While we follow the BLSTM architectures of Open-Unmix  \cite{Stoter2019} 
and RNNDownBeatProcessor
\cite{Bock2016d} 
for the first two modules of our model, the parameters of these two modules, as well as those of the third fuser module, are all to be learned from the training set. 
Specifically, the loss of all these three modules are detached so that the parameters of each module are optimized using its own training loss function. 
All the three beat trackers (i.e. mixture, drum, non-drum) are independently supervised by the same set of beat/downbeat annotations for each song in the training set, in the light that the corresponding audio signals share the same metrical structure. 

For the last module, we use the HMM available in Madmom  \cite{Bock2016}, with the hyperparameters tuned using our validation set.

\vspace{-1mm}
\subsection{Feature Separation}
\label{sec:methodology:separation}
\noindent The input audio is firstly represented by the combination of three magnitude spectrograms and their first-order derivatives with different window sizes following \cite{Bock2016d}. This representation, dubbed the mixture feature in Fig. \ref{fig:sys_overview}, is then fed to a drum/non-drum source separation module with a three-layer BLSTM architecture akin to the Open-Umix (OU) model \cite{Stoter2019} to generate the drum and non-drum features by masking. 
The feature separation module is supervised by the mean squared error  between the features it produced and the features calculated from the reference drum and non-drum audio computed in advance by another 
source separation library named Spleeter \cite{Hennequin2019}. The officially pretrained Spleeter was trained on a large private dataset and was claimed to outperform  the officially pretrained OU. 
We use the officially pretrained Spleeter for preparing the supervisory drum and non-drum audio signals, instead of adopting its fully-convolutional model architecture for our own feature separation module, as the recurrent neural network based architecture adopted by OU can more easily deal with variable-length audio signals at the inference time. 



\vspace{-1mm}
\subsection{Parallel Beat/Downbeat Trackers}
\noindent It happens that the RNNDownBeatProcessor of Madmom also has a three-layer BLSTM architecture \cite{Bock2016}.
We implement such an architecture on our own following the referenced publication \cite{Bock2016d}. 
We use the same BLSTM architecture for our three  trackers, which take as input the mixture feature, drum feature, and non-drum feature respectively. 
They each produces a $3\times T$ matrix indicating the activation likelihood of beat, downbeat, and ``non-beat'' of each of the $T$ time frames, shown as separate rows in each $3\times 8$ matrix in Fig. \ref{fig:sys_overview} ($T=8$ here).
Namely, each tracker tracks the presence of beats and downbeats jointly (and with separate activation functions), regarding the rest as the non-beat.
Cross-entropy loss is applied with an empirically determined weight (67:200:1) to deal with the class imbalance among beat, downbeat, and non-beat.

\vspace{-1mm}
\subsection{Fusion Mechanism}
\noindent At the fusion stage, the activation functions from the parallel trackers are combined and used as input to the \emph{fuser tracker}, 
also a three-layer BLTSM but with smaller input feature size and less hidden units (10 units in our implementation).
The output of the fuser is also a $3\times T$ matrix.
We implement two variants of the fuser. The first variant, \textbf{DA1} (drum-aware ensemble), simply concatenates the output of the three individual trackers as its input.
The second variant, \textbf{DA2} (two-stage DA), uses additionally the output of the last BLSTM layer of the preceding trackers as input, with an enlarged number of 25 hidden units in the fuser tracker. We assume DA2 may work better as it has access to not only the final output but also the intermediate states of the individual trackers.


\begin{table*}[th]
\caption{Evaluation result (in F1 score) on the test sets; the three best result in each column underlined, and the best one italicized. Daggers (or two daggers) denote that the model have partially (or completely) seen the specific test sets. 
Bolded ones indicate the one-tailed t-test result for the performance advantage over Baseline is significant (p-valuee$<$0.05).}

\centering
\begin{tabular}{ll|llll|c|llll|c}
\toprule
\multirow{2}{*}{\textbf{Model}}  & 
&
\multicolumn{5}{c|}{\textbf{Beat}}
& \multicolumn{5}{c}{\textbf{Downbeat}}
\\
\cline{3-7}
\cline{8-12}
  &      & HJDB  & Merged & Rock           & ASAP  & Mean  & HJDB     & Merged & Rock           & ASAP  & Mean  \\
\midrule
\multicolumn{2}{l|}{Baseline (our implementation of \cite{Bock2016d})}   & 0.886 & \underline{0.884}  & \underline{0.908}          & 0.585 & 0.762 & 0.699    & 0.715  & 0.842          & 0.399 & 0.599 \\
\multicolumn{2}{l|}{Madmom API \cite{Bock2016d}}   & 0.995\footnote[8] & 0.904\footnote[2] & 0.957\footnote[8] & 0.468 & 0.746 & 0.970\footnote[8] & 0.733\footnote[2] & 0.927\footnote[8] & 0.275 & 0.616 \\
\midrule
DA1       & mix    & 0.875 & 0.876  & 0.901          & 0.584 & 0.756 & 0.656    & 0.713  & 0.817          & 0.396 & 0.585 \\
DA1       & drum   & 0.897 & 0.847  & 0.891          & 0.512 & 0.722 & 0.720    & 0.669  & 0.768          & 0.273 & 0.528 \\
DA1       & nodrum & 0.759 & 0.854  & 0.878          & 0.578 & 0.724 & 0.487    & 0.671  & 0.814          & 0.393 & 0.542 \\
\cdashline{2-2}[1pt/1pt]
DA1       & bagging  & 0.875 & 0.875  & 0.907 & 0.574 & 0.753 & 0.754    & 0.736  & \underline{0.844} & 0.383 & 0.608 \\
DA1 &
  fuser &
  \textbf{\underline{0.914}} &
  \textbf{{\emph{\underline{0.894}}}}&
  0.907 &
  \underline{0.587} &
  \underline{0.770} &
  \textbf{\underline{0.769}} &
  \textbf{\underline{0.743}} &
  0.842 &
  \underline{0.402} &
  \underline{0.620} \\
\midrule
DA2      & mix    & 0.879 & 0.877  & 0.901          & \underline{0.587} & 0.759 & 0.651    & 0.701  & 0.831          & \underline{0.403} & 0.587 \\
DA2      & drum   & \underline{0.906} & 0.847  & 0.886          & 0.522 & 0.727 & 0.750    & 0.666  & 0.775          & 0.273 & 0.534 \\
DA2      & nodrum & 0.775 & 0.863  & 0.885          & 0.584 & 0.732 & 0.521    & 0.699  & 0.813          & 0.402 & 0.558\\
\cdashline{2-2}[1pt/1pt]
DA2 &
  bagging &
  0.904 &
  0.880 &
  \underline{0.910} &
  0.582 &
  \underline{0.764} &
  {\emph{\underline{0.802}}} &
  \underline{0.739} &
  \underline{0.853} &
  0.387 &
  \underline{0.621} \\
DA2 &
  fuser &
  \textbf{{\emph{\underline{0.914}}}} &
  \textbf{\underline{0.891}} &
  {\emph{\underline{0.911}}} &
  \textbf{{\emph{\underline{0.596}}}} &
  {\emph{\underline{0.774}}} &
  \textbf{\underline{0.775}} &
  \textbf{\emph{\underline{0.743}}} &
  \textbf{\emph{\underline{0.853}}} &
  \textbf{{\emph{\underline{0.415}}}} &
  {\emph{\underline{0.628}}} \\
\bottomrule
\end{tabular}
\label{tab:result_all}
\end{table*}

\section{Experimental Setup}
\label{sec:exp}

\subsection{Datasets}
\label{sec:exp:data}

\noindent Table \ref{tab:dataset_overview} shows the datasets adopted in our experiment. We use the first ten datasets to train our models. Specifically, each of these datasets is randomly split into 80\%, 10\%, 10\% to become part of the training, validation, and the first test set called ``Merged.''
The last three datasets are kept as completely unseen test sets, including ASAP---a drum-less collection of classical piano music \cite{asap-dataset}, Rock---200 of the Rolling Stone magazine’s list of the ``500 Greatest Songs of All Time'' \cite{rockdataset}, and HJDB---a drum-heavy dataset comprising Hardcore, Jungle and Drum\&Bass music excerpts \cite{Hockman2012}.

To have an idea of the difference in drum usage in these datasets, we also present in Table \ref{tab:dataset_overview} 
the ``drum presence'' rate (in \%) calculated by the following ad-hoc method. 
For each song, we employ the officially pretrained 
Spleeter \cite{Hennequin2019} to derive its drum stem. 
Then, we compute the mean value of the drum stem's absolute magnitude (ABSM) in the time domain, and consider the song as a drum-less song if the ABSM is less than 0.01, an empirically set threshold.  The drum presence rate is defined as the percetage of songs in a dataset that are not drum-less. 
We can see that the datasets differ a lot in this rate.

\vspace{-1mm}
\subsection{Baselines}

\noindent We implement the following baselines or ablated variants to validate the effectiveness of the proposed ensemble architecture, which is denoted as \textbf{fuser} in Table \ref{tab:result_all}.
\begin{itemize}
    \item \textbf{baseline}: The widely-adopted architecture comprising a single BLSTM-based beat/downbeat tracker and an HMM postprocessing module, implemented following Madmom \cite{Bock2016d}. Namely, neither source separation nor fusion is used. Similar to \cite{Krebs2016}, we find increasing the layers of BLSTM not helpful and retain the three-layer BLSTM setting. 
    \item \textbf{mix}, \textbf{drum}, \textbf{nodrum}: The \emph{single-headed} cases where we train the proposed model as usual yet use the output of only one of the individual trackers (i.e., mixture tracker, drum tracker, or non-drum tracker, respectively) as input to the HMM to get the final estimate at evaluation time. This is to validate the advantage of fusion.
    \item \textbf{bagging}: We implement a simple fuser that takes the averaged activation of the three individual trackers, instead of using the sophisticated BLSTM-based fusion. 
\end{itemize}




\subsection{Model Training \& Evaluation Metrics}
\noindent  Our models are trained with the Lookahead Adam optimizer \cite{NEURIPS2019_90fd4f88} with \num{e-2} learning rate. If no improvement on validation loss can be observed for 20 epochs, we 
reduce the learning rate by a factor of five. To reduce the influence of random initialization, we repeat the training of all models for five times and report the averaged evaluation results. 
Following the convention of beat/downbeat tracking (e.g., \cite{Bock}), we report the F-measure (F1) with a tolerance window of $\pm 70$~ms \cite{Raffel14mir_eval:a}.

\section{Results and Discussion}
\label{sec:result}


Table \ref{tab:result_all} shows the evaluation result. Looking at the first row shows that the  Madmom-like \emph{baseline} already performs quite well across the test sets, especially for beat tracking. The F1 scores degrade and exhibit larger variation across the datasets for downbeat tracking, suggesting downbeats are in general more difficult to track than beats. Both tasks are challenging on  ASAP, which is expected as our training set 
does not contain many classical music pieces.

The next block of rows in Table \ref{tab:result_all}, namely the result of the DA1 based models, shows that the proposed ensemble model \emph{DA1\_fuser} consistently outperforms its ablated versions across different test sets in both beat \& downbeat tracking, validating the effectiveness of the proposed design. Compared to the result of the \emph{baseline}, we see salient performance gain in beat and downbeat tracking for HJDB (+7.0\% relative improvement in downbeat), which comprises songs from musical genres that are seldom seen in the training set. 
Looking at the performance of the individual trackers \emph{DA1\_mix}, \emph{DA1\_
drum}, \emph{DA1\_nodrum} for HJDB shows that the performance of \emph{DA1\_drum} is the strongest among the three, which is again expected since HJDB features fairly high drum presence rate. 
This might also explain why the fused result  outperforms the \emph{baseline}. We take this as an evidence of the benefit of building tailored trackers for percussive and non-percussive sounds to take care of the versatile role of the drums in different music signals.

Table \ref{tab:result_all} also shows that \emph{DA1\_drum} performs the worst among the three individual trackers for beat \& downbeat tracking for ASAP. While the simple fusion method \emph{DA1\_bagging} may suffer from the inconsistent performance of the individual trackers across datasets, \emph{DA1\_fuser} can nicely aggregate the result of the three trackers.
Instead of simply take the average, \emph{DA1\_fuser} might be aware of the absence of the drum sounds from the output of the drum tracker and accordingly decide to rely more on the other two individual trackers.

The last block of rows in Table \ref{tab:result_all}, the result of the DA2 based models, shows that DA2 is in general slightly better thanDA1. \emph{DA2\_fuser} outperforms \emph{DA1\_fuser} in most cases and leads to the best mean F1 scores in both beat tracking (0.774) and downbeat tracking (0.628). This indicates that we can achieve more effective fusion by taking advantage of the intermediate result of the individual trackers.


In mean F1 scores, \emph{DA2\_fuser} outperforms the \emph{baseline} by 0.012 in tracking beats and by 0.029 in downbeats. The larger improvement in the latter might be partly due to the available room for improvement, but might also suggest the prediction of downbeats benefits more from separating the effect of the percussive and non-percussive instruments. 

One-tailed 
$t$-test shows that the per-song performance difference in F1 (averaged over the five runs) between either the pair \emph{baseline} vs. \emph{DA1\_fuser}, or  \emph{baseline} vs. \emph{DA2\_fuser}, is 
significant ($p$-value$<$0.05) in many cases (bolded in Table \ref{tab:result_all}).
We also see significant performance difference between \emph{DA1\_fuser}, \emph{DA2\_fuser} for both beats \& downbeats on ASAP.

Lastly, we evaluate the  models using another metric, called CMLt \cite{beatles}, defined as the ratio of the total number of correct beats (or downbeats) at correct metrical level to the total number of annotations. While  F1 treats a beat/downbeat estimate as correct by whether or not it falls within a fixed-length tolerance window ($\pm 70$~ms), CMLt adopts a variable-length tolerance window (i.e., $\pm 17.5$\% of the current inter-annotation interval), and additionally  
demands
consistency between the inter-annotation interval and the inter-beat (or -downbeat) interval.
This downplays 
the effect of tempo drift where occasional beats will be in phase \cite{beatles}. 
Table \ref{tab:continuity} shows that both \emph{DA1\_fuser} and \emph{DA2\_fuser} still outperform the \emph{baseline} in CMLt. 
In many cases (highlighted in Table \ref{tab:continuity}) the per-song performance difference in CMLt between either  \emph{baseline} vs. \emph{DA1\_fuser}, or  \emph{baseline} vs. \emph{DA2\_fuser}, is also significant ($p$-value$<$0.05) under the one-tailed $t$-test.

\begin{table}
\caption{CMLt results of the Baseline (BSL), DA\_fuser and DA2\_fuser}
\centering
\begin{tabular}{l|cccc}
\toprule
& \multicolumn{4}{c}{\textbf{Beat}~/~\textbf{Downbeat}}        \\
 & HJDB & Merged & Rock & ASAP \\
\midrule
BSL &
  0.870~/~0.748 &
  0.784~/~0.638 &
  0.795~/~0.750 &
  0.296~/~0.217 \\
DA1 &
  \textbf{0.902}~/~\textbf{0.804} &
  \textbf{0.805}~/~\textbf{0.668} &
  0.807~/~0.761 &
  0.293~/~0.212 \\
DA2 &
  \textbf{0.904}~/~\textbf{0.811} &
  \textbf{0.798}~/~\textbf{0.669} &
  \textbf{0.805}~/~\textbf{0.767} &
  \textbf{0.307}~/~\textbf{0.225} \\
\bottomrule
\end{tabular}
\label{tab:continuity}
\end{table}

\section{Conclusion}

Aiming to improve the performance of joint beat/downbeat tracking across different types of musical audio signals, we have presented in this paper a drum-aware ensemble architecture that employs and fuses the result of multiple parallel beat/downbeat trackers for different sound sources in an input signal. Experiment results demonstrated the advantage of such a new multi-model approach over a single-model, Madmom's RNNDownBeatProcessor-like baseline \cite{Bock2016d}. Ablation experiments also confirmed the effectiveness of the proposed BLSTM-based fusion mechanism. 
For future work, we are interested in experimenting with different architectures for the tracking and fusion modules, such as  temporal convolutional networks \cite{Davies2019} or Transformer \cite{vaswani2017attention},
and in further improving the performance of beat/downbeat tracking for classical music.

\begin{table*}[th]
\caption{Evaluation result of SDA models (in F1 score) on the test sets; 
Bolded ones indicate the best head of each model on that column.}
\centering
\begin{tabular}{ll|llll|c|llll|c}
\toprule
\multirow{2}{*}{\textbf{Model}}  & 
&
\multicolumn{5}{c|}{\textbf{Beat}}
& \multicolumn{5}{c}{\textbf{Downbeat}}        
\\
\cline{3-7}
\cline{8-12}
&                      & HJDB           & Merged         & Rock           & ASAP           & Mean           & HJDB           & Merged         & Rock           & ASAP           & Mean           \\
\midrule
SDA1               & mix                   & 0.885          & 0.885          & 0.917          & \textbf{0.587} & 0.764          & 0.661          & 0.707          & 0.838          & 0.404          & 0.592          \\
SDA1               & drum                  & 0.897          & 0.781          & 0.871          & 0.130          & 0.543          & 0.763          & 0.596          & 0.724          & 0.043          & 0.415          \\
SDA1               & nodrum                & 0.701          & 0.835          & 0.875          & 0.572          & 0.705          & 0.421          & 0.666          & 0.801          & 0.395          & 0.527          \\
\cdashline{2-2}[1pt/1pt]
SDA1               & bagging               & 0.846          & 0.845          & \textbf{0.919} & 0.452          & 0.691          & 0.758          & 0.708          & \textbf{0.850} & 0.291          & 0.564          \\
SDA1               & fuser                 & \textbf{0.927} & \textbf{0.894} & 0.916          & 0.585          & \textbf{0.773} & \textbf{0.800} & \textbf{0.735} & 0.844          & \textbf{0.414} & \textbf{0.630} \\

\midrule
SDA2               & mix                   & 0.882          & 0.871          & 0.888          & \textbf{0.596} & 0.760          & 0.701          & 0.707          & 0.825          & 0.428          & 0.607          \\
SDA2               & drum                  & 0.886          & 0.793          & 0.858          & 0.173          & 0.559          & 0.745          & 0.606          & 0.698          & 0.057          & 0.416          \\
SDA2               & nodrum                & 0.679          & 0.846          & 0.871          & 0.577          & 0.705          & 0.391          & 0.673          & 0.795          & 0.402          & 0.525          \\
\cdashline{2-2}[1pt/1pt]
SDA2               & bagging               & 0.855          & 0.855          & \textbf{0.896} & 0.457          & 0.694          & 0.775          & 0.727          & \textbf{0.835} & 0.313          & 0.579          \\
SDA2               & fuser                 & \textbf{0.908} & \textbf{0.890} & 0.895          & 0.587          & \textbf{0.767} & \textbf{0.808} & \textbf{0.744} & 0.817          & \textbf{0.429} & \textbf{0.635}
\\
\bottomrule
\end{tabular}
\label{tab:result_sda}
\end{table*}

\section*{Appendix}
We add to this ArXiv preprint some materials that cannot fit in the SPL paper due to the space limit. First, we comment on the difference between the present work and our earlier work adopting source separation for data augmentation (Aug4Beat) \cite{chiu21eusipco}. Second, we present the result of an ablated version of our system which uses the officially pretrained Spleeter module instead of a trainable feature separation module. Lastly, we discuss the averaged activation functions of different trackers on two test sets.

\subsection{On the Difference between Aug4Beat and the Present Work}
Aug4Beat~\cite{chiu21eusipco}, a precursor of the present paper, represents our first attempt in using source separation to improve the performance of beat and downbeat tracking.
The main idea of Aug4Beat is to use source separation as a way to create additional data for model training. 
We consider similar experimental setup there and here, using exactly the same training sets and almost the same test sets, making it valid to compare the results in these two papers. But, we remark that we actually use a different optimizer in this paper, and this may play some role. 
Specifically, in Aug4Beat, we adopt stochastic gradient descent with momentum (SGDM) with a learning rate of \num{e-3}, following \cite{Bock2016d}. However, in this paper we adopt the Lookahead Adam optimizer \cite{NEURIPS2019_90fd4f88} with \num{e-2} learning rate, following \cite{Bock}. We find the new optimizer greatly reduces  training time and achieves better overall performance for the  DA architecture. 
Comparing the result of the `baseline' systems in Table III of~\cite{chiu21eusipco} and Table II of this paper, we see that SGDM performs better than Lookahead Adam  on ASAP, but much worse on the Rock dataset. 
Future work can be done to more closely examine the influence of the optimizers, and to integrate the idea of data augmentation and DA. 

\subsection{On the Effect of the Feature Separation Module}

We opt for training the feature separation module together with other modules 
(cf. Section \ref{sec:methodology:separation})
as STOA beat tracking models and source separation (SS) models tend to use different input features; the former usually use magnitude-filtered spectrograms plus their first-order derivatives (e.g., \cite{Bock2016d}), while the latter use high resolution STFT spectrograms (e.g., Spleeter \cite{Hennequin2019}).
Accordingly, adopting the pretrained Spleeter ``as is'' in our system would require extra feature processing to convert the output of the source separation model to become the input to the beat and downbeat tracking system. 

To empirically examine the effect of this, we implement such a ``Spleeter DA'' (or SDA for short) where we replace the trainable feature separation module with a fixed, officially pretrained Spleeter model.
Table \ref{tab:result_sda} shows the result, averaged over five runs for each model. We can see that, except for the beat F-score on ASAP, the \emph{SDA\_fusers} benefit from the DA idea just like the proposed DA models reported in Table \ref{tab:result_all}. Similar trends between SDA and DA models can also be observed. For example, \emph{fusers} gain more improvements on HJDB and Merged test sets and \emph{bagging} performs well on the Rock test set. The major differences between SDA and DA are in the results on ASAP. First, unlike the \emph{DA\_fusers}, the \emph{SDA\_fusers} do not gain improvement on ASAP beat performance. Second, the \emph{SDA\_drums} get significantly worse performance on ASAP. We conjecture that this is due to the high-quality SS results of Spleeter. As Spleeter could perform very well on SS, the drum stems it isolates out from pieces in ASAP would be nearly silent, providing  \emph{SDA\_drum} nearly no information to track the beats and downbeats. In similar lights, we can also see that the \emph{SDA\_nodrums} perform much worse than the \emph{DA\_nodrums} on HJDB. 
We take this as an empirical support of the design proposed in Section \ref{sec:methodology:separation}.

\subsection{Inside the Activation Functions}
\begin{figure}
\centerline{\includegraphics[width=\columnwidth]{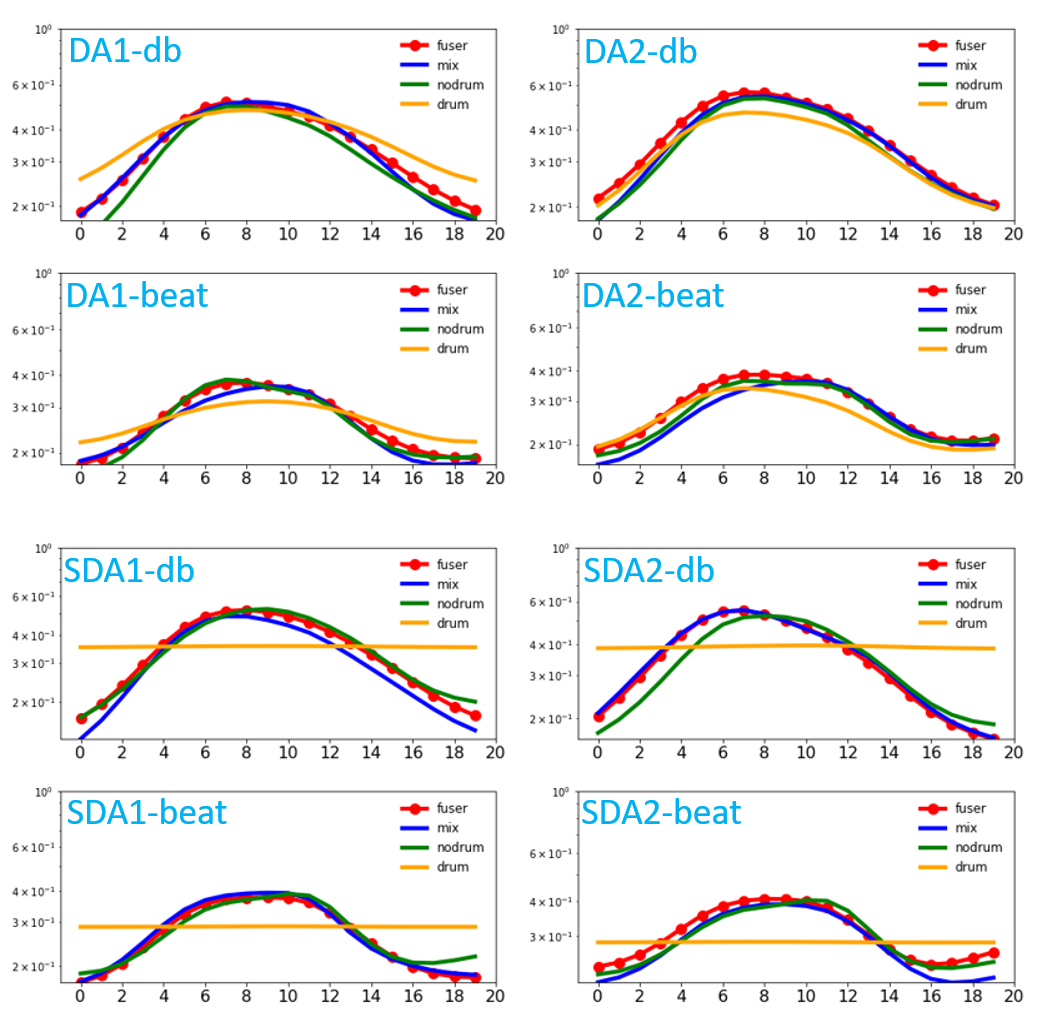}}
\caption{Averaged activation function derived on ASAP test set for regions from 10 frames before beat or downbeat position to 10 frames after beat or downbeat position. `db' denotes downbeat.}
\label{fig:asap}
\end{figure}

\begin{figure}
\centerline{\includegraphics[width=\columnwidth]{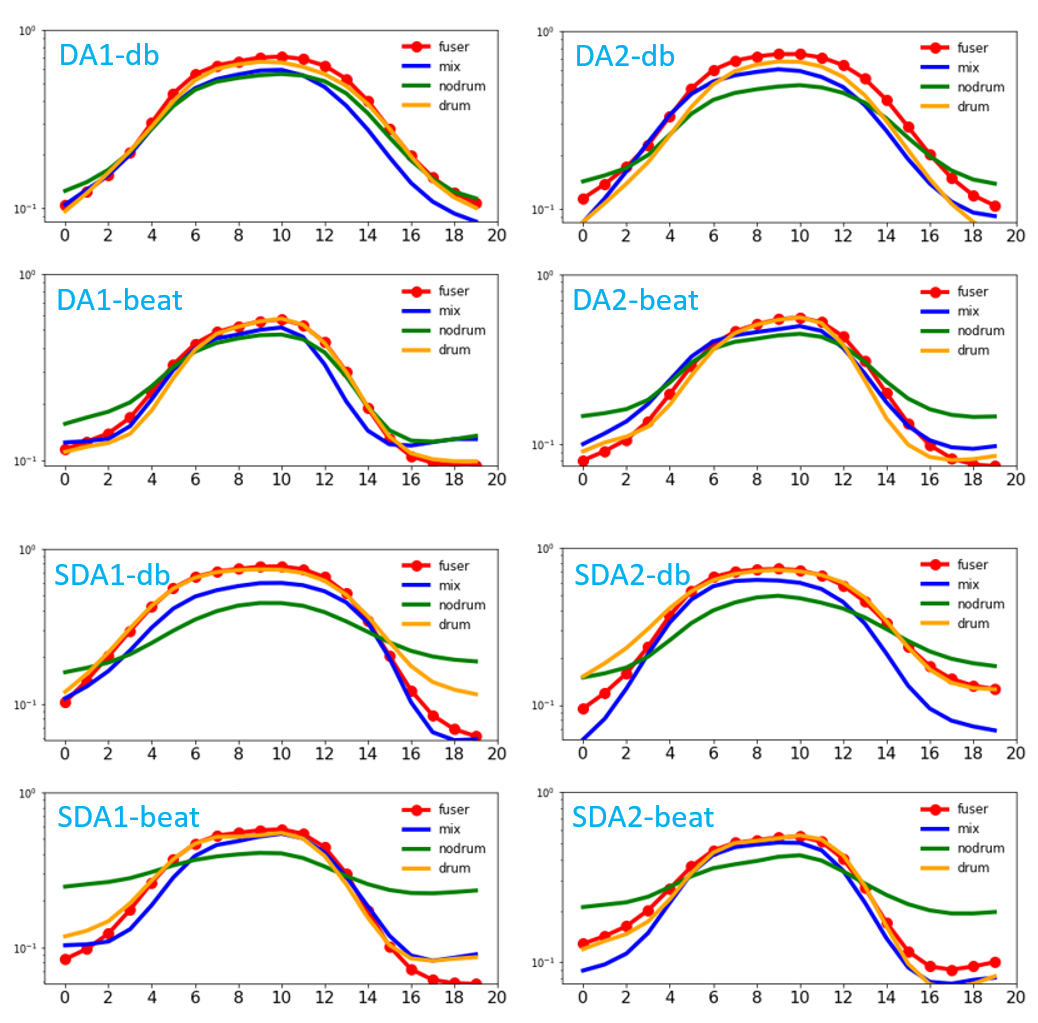}}
\caption{Averaged activation function derived on HJDB test set for regions from 10 frames before beat or downbeat position to 10 frames after beat or downbeat position.}

\label{fig:hjdb}
\end{figure}

To gain insights into the behavior of the DA idea, we plot the ``averaged'' activation function of different trackers for the regions from 10 frames before beat/downbeat position to 10 frames after beat/downbeat position, over all the pieces in a specific test set. This approach enables us to see how each head (i.e. \emph{fuser}, \emph{mix}, \emph{drum}, \emph{nondrum}) reacts to a beat/downbeat on average. Figures \ref{fig:asap} and \ref{fig:hjdb} show the results for ASAP and HJDB, respectively.
We can see that for both test sets, each head reacts differently (and sometimes complementarily) to a beat/downbeat. And the \emph{fusers} of both SDA and DA generally integrate information from other heads and produce the highest peak near beat/downbeat position (i.e. position 10.0 at x-axis).
More specifically, we can observe that the \emph{drum} heads of DA and SDA are clearly more confident than the \emph{mix} heads on HJDB dataset. And on ASAP dataset, the \emph{nodrum} heads still react differently than the \emph{mix} heads. These observations echo our hypothesis that the models are driven by the sound source composition of training data. As a \emph{mix} head is trained with mixture input feature, it learns to rely on patterns related to both drum and non-drum feature. Therefore, it could be less sensitive than a \emph{drum} head to specific drum-related patterns. On the other hand, since the \emph{nodrum} head is trained with non-drum features, it would be less sensitive than the \emph{mix} head to the drum-related patterns. In summary, the source-separation design of the system enables the individual trackers (i.e. \emph{mix}, \emph{drum}, \emph{nodrum}) to learn to interpret the input feature differently. And, the \emph{fuser} head therefore has higher chance to access to more or better preserved beat/downbeat related information/pattern and generate more confident results.



\bibliographystyle{IEEEtran}
\bibliography{spl2021}

\begin{thebibliography}{10}
\providecommand{\url}[1]{#1}
\csname url@samestyle\endcsname
\providecommand{\newblock}{\relax}
\providecommand{\bibinfo}[2]{#2}
\providecommand{\BIBentrySTDinterwordspacing}{\spaceskip=0pt\relax}
\providecommand{\BIBentryALTinterwordstretchfactor}{4}
\providecommand{\BIBentryALTinterwordspacing}{\spaceskip=\fontdimen2\font plus
\BIBentryALTinterwordstretchfactor\fontdimen3\font minus
  \fontdimen4\font\relax}
\providecommand{\BIBforeignlanguage}[2]{{%
\expandafter\ifx\csname l@#1\endcsname\relax
\typeout{** WARNING: IEEEtran.bst: No hyphenation pattern has been}%
\typeout{** loaded for the language `#1'. Using the pattern for}%
\typeout{** the default language instead.}%
\else
\language=\csname l@#1\endcsname
\fi
#2}}
\providecommand{\BIBdecl}{\relax}
\BIBdecl

\bibitem{ballroom2}
F.~Krebs, S.~B{\"o}ck, and G.~Widmer, ``Rhythmic pattern modeling for beat and
  downbeat tracking in musical audio,'' in \emph{Proc. Int. Soc. Music Inf.
  Retr. Conf.}, 2013.

\bibitem{Bock2014b}
S.~B{\"{o}}ck, F.~Krebs, and G.~Widmer, ``{A multi-model approach to beat
  tracking considering heterogeneous music styles},'' \emph{Proc. Int. Soc.
  Music Inf. Retr. Conf.}, pp. 603--608, 2014.

\bibitem{davies11spl}
M.~E.~P. {Davies}, N.~{Degara}, and M.~D. {Plumbley}, ``Measuring the
  performance of beat tracking algorithms using a beat error histogram,''
  \emph{IEEE Signal Processing Letters}, vol.~18, no.~3, pp. 157--160, 2011.

\bibitem{Holzapfel2012b}
A.~Holzapfel, M.~E.~P. Davies, J.~R. Zapata, J.~L. Oliveira, and F.~Gouyon,
  ``Selective sampling for beat tracking evaluation,'' \emph{IEEE Trans. Audio,
  Speech Lang. Process.}, vol.~20, no.~9, pp. 2539--2548, 2012.

\bibitem{durand17taslp}
S.~{Durand}, J.~P. {Bello}, B.~{David}, and G.~{Richard}, ``Robust downbeat
  tracking using an ensemble of convolutional networks,'' \emph{IEEE/ACM Trans.
  Audio, Speech, and Language Processing}, vol.~25, no.~1, pp. 76--89, 2017.

\bibitem{benetos19spm}
E.~{Benetos} \emph{et~al.}, ``Automatic music transcription: An overview,''
  \emph{IEEE Signal Processing Magazine}, vol.~36, no.~1, pp. 20--30, 2019.

\bibitem{burloiu16smc}
G.~Burloiu, ``An online tempo tracker for automatic accompaniment based on
  audio-to-audio alignment and beat tracking,'' in \emph{Proc. Sound and Music
  Computing Conf.}, 2016, pp. 93--98.

\bibitem{huang20mm}
Y.-S. Huang and Y.-H. Yang, ``{Pop Music Transformer}: Beat-based modeling and
  generation of expressive pop piano compositions,'' in \emph{Proc. ACM Int.
  Conf. Multimedia}, 2020, p. 1180–1188.

\bibitem{Bock2016d}
S.~B{\"{o}}ck, F.~Krebs, and G.~Widmer, ``{Joint beat and downbeat tracking
  with recurrent neural networks},'' \emph{Proc. Int. Soc. Music Inf. Retr.
  Conf.}, pp. 255--261, 2016.

\bibitem{Bock2016}
S.~B{\"{o}}ck, F.~Korzeniowski, J.~Schl{\"{u}}ter, F.~Krebs, and G.~Widmer,
  ``{Madmom: A new Python audio and music signal processing library},''
  \emph{Proc. ACM Multimed. Conf.}, pp. 1174--1178, 2016.

\bibitem{Fuentes2018}
M.~Fuentes, B.~McFee, H.~C. Crayencour, S.~Essid, and J.~P. Bello, ``Analysis
  of common design choices in deep learning systems for downbeat tracking,'' in
  \emph{Proc. Int. Soc. Music Inf. Retr. Conf.}, 2018, pp. 106--112.

\bibitem{Bock}
S.~B{\"{o}}ck and M.~E.~P. Davies, ``{Deconstruct, analyse, reconstruct: How to
  improve tempo, beat, and downbeat estimation},'' in \emph{Proc. Int. Soc.
  Music Inf. Retr. Conf.}, 2020, p. 574–582.

\bibitem{giorgi20ismir}
B.~D. Giorgi, M.~Mauch, and M.~Levy, ``Downbeat tracking with tempo invariant
  convolutional neural networks,'' in \emph{Proc. Int. Soc. Music Inf. Retr.
  Conf.}, 2020, p. 216–222.

\bibitem{cano20ismir}
E.~Cano, F.~M. Angel, G.~A.~L. Gil, J.~R. Zapata, A.~Escamilla, J.~F.~A.
  Londoño, and M.~B. Pelaez, ``Sesquialtera in the {Colombian Bambuco}:
  Perception and estimation of beat and meter,'' in \emph{Proc. Int. Soc. Music
  Inf. Retr. Conf.}, 2020, p. 409–415.

\bibitem{pedersoli20ismir}
F.~Pedersoli and M.~Goto, ``Dance beat tracking from visual information
  alone,'' in \emph{Proc. Int. Soc. Music Inf. Retr. Conf.}, 2020, p.
  400–408.

\bibitem{Stoter2019}
F.-R. St{\"{o}}ter, S.~Uhlich, A.~Liutkus, and Y.~Mitsufuji, ``{Open-Unmix - A
  reference implementation for music source separation},'' \emph{J. Open Source
  Softw.}, vol.~4, no.~41, p. 1667, 2019.

\bibitem{Hennequin2019}
R.~Hennequin, F.~V. A.~Khlif, and M.~Moussallam, ``{Spleeter: A fast and
  state-of-the art music source separation tool with pre-trained models},''
  \emph{J. Open Source Softw.}, vol.~5, no.~50, p. 2154, 2020.

\bibitem{liu19ijcai}
J.-Y. Liu and Y.-H. Yang, ``Dilated convolution with dilated {GRU} for music
  source separation,'' in \emph{Proc. Int. Joint Conf. Artificial
  Intelligence}, 2019, pp. 4718--4724.

\bibitem{chiu20mmsp}
C.-Y. Chiu, W.-Y. Hsiao, Y.-C. Yeh, Y.-H. Yang, and A.~W.~Y. Su,
  ``Mixing-specific data augmentation techniques for improved blind
  violin/piano source separation,'' in \emph{Proc. IEEE Int. Workshop on
  Multimedia Signal Processing}, 2020.

\bibitem{Gouyon_acoustic_cue}
F.~{Gouyon}, G.~{Widmer}, X.~{Serra}, and A.~{Flexer}, ``{Acoustic cues to beat
  induction: A machine learning perspective},'' \emph{Music Perception},
  vol.~24, no.~2, pp. 177--188, 2006.

\bibitem{Jia2019}
B.~Jia, J.~Lv, and D.~Liu, ``{Deep learning-based automatic downbeat tracking:
  a brief review},'' \emph{Multimedia Systems}, vol.~25, no.~6, pp. 617--638,
  2019.

\bibitem{Bock2019}
S.~B{\"{o}}ck, M.~E.~P. Davies, and P.~Knees, ``{Multi-task learning of tempo
  and beat: Learning one to improve the other},'' \emph{Proc. Int. Soc. Music
  Inf. Retr. Conf.}, pp. 486--493, 2019.

\bibitem{Grosche2010a}
P.~Grosche, M.~M{\"{u}}ller, and C.~S. Sapp, ``What makes beat tracking
  difficult? a case study on {Chopin Mazurkas},'' \emph{Proc. Int. Soc. Music
  Inf. Retr. Conf.}, pp. 649--654, 2010.

\bibitem{asap-dataset}
F.~Foscarin, A.~McLeod, P.~Rigaux, F.~Jacquemard, and M.~Sakai, ``{ASAP}: a
  dataset of aligned scores and performances for piano transcription,'' in
  \emph{Proc. Int. Soc. Music Inf. Retr. Conf.}, 2020, pp. 534--541.

\bibitem{Zapata2013a}
J.~R. Zapata and E.~Gomez, ``{Using voice suppression algorithms to improve
  beat tracking in the presence of highly predominant vocals},'' \emph{Proc.
  IEEE Int. Conf. Acoust. Speech Signal Process.}, pp. 51--55, 2013.

\bibitem{Goto97musicunderstanding}
M.~Goto and Y.~Muraoka, ``{Music understanding at the beat level real-time beat
  tracking for audio signals},'' \emph{Proc. Int. Joint Conf. Artificial
  Intelligence Workshop on Computational Auditory Scene Analysis}, 1995.

\bibitem{GOTO1999311}
------, ``Real-time beat tracking for drumless audio signals: Chord change
  detection for musical decisions,'' \emph{Speech Communication}, vol.~27,
  no.~3, pp. 311--335, 1999.

\bibitem{Goto2001}
M.~Goto, ``{An audio-based real-time beat tracking system for music with or
  without drum-sounds},'' \emph{Journal of New Music Research}, vol.~30, no.~2,
  pp. 159--171, 2001.

\bibitem{McFee2014_onsetAGG}
B.~McFee and D.~P. Ellis, ``Better beat tracking through robust onset
  aggregation,'' in \emph{Proc. IEEE Int. Conf. Acoust. Speech Signal
  Process.}, 2014, pp. 2154--2158.

\bibitem{percussivebeat2013}
A.~Robertson, A.~Stark, and M.~E.~P. Davies, ``Percussive beat tracking using
  real-time median filtering,'' \emph{Int. Workshop on Machine Learning and
  Music}, 2013.

\bibitem{Durand2014}
S.~Durand, B.~David, and G.~Richard, ``{Enhancing downbeat detection when
  facing different music styles},'' in \emph{Proc. IEEE Int. Conf. Acoust.
  Speech Signal Process.}, 2014, pp. 3132--3136.

\bibitem{Durand2015a}
S.~Durand, J.~P. Bello, B.~David, and G.~Richard, ``{Downbeat tracking with
  multiple features and deep neural networks},'' \emph{Proc. IEEE Int. Conf.
  Acoust. Speech Signal Process.}, pp. 409--413, 2015.

\bibitem{Durand2017a}
------, ``{Robust downbeat tracking using an ensemble of convolutional
  networks},'' \emph{IEEE/ACM Trans. Audio Speech Lang. Process.}, vol.~25,
  no.~1, pp. 72--85, 2017.

\bibitem{Krebs2016}
F.~Krebs, S.~B{\"{o}}ck, M.~Dorfer, and G.~Widmer, ``{Downbeat tracking using
  beat-synchronous features and recurrent neural networks},'' \emph{Proc. Int.
  Soc. Music Inf. Retr. Conf.}, pp. 129--135, 2016.

\bibitem{Schreiber2018}
H.~Schreiber and M.~M{\"{u}}ller, ``{A single-step approach to musical tempo
  estimation using a convolutional neural network},'' \emph{Proc. Int. Soc.
  Music Inf. Retr. Conf.}, pp. 98--105, 2018.

\bibitem{rwcdatabase}
M.~Goto, H.~Hashiguchi, T.~Nishimura, and R.~Oka, ``{RWC Music Database}:
  {Popular, Classical, and Jazz} music databases,'' in \emph{Proc. Int. Soc.
  Music Inf. Retr. Conf.}, 2002, pp. 287--288.

\bibitem{rwc_musicgenre}
------, ``{RWC Music Database: Music genre database and musical instrument
  sound database.}'' in \emph{Proc. Int. Soc. Music Inf. Retr. Conf.}, 2003,
  pp. 229--230.

\bibitem{ballroom1}
F.~{Gouyon}, A.~{Klapuri}, S.~{Dixon}, M.~{Alonso}, G.~{Tzanetakis}, C.~{Uhle},
  and P.~{Cano}, ``An experimental comparison of audio tempo induction
  algorithms,'' \emph{IEEE Transactions on Audio, Speech, and Language
  Processing}, vol.~14, no.~5, pp. 1832--1844, 2006.

\bibitem{hainsworth}
M.~Macleod and S.~Hainsworth, ``Particle filtering applied to musical tempo
  tracking,'' \emph{EURASIP Journal on Advances in Signal Processing}, vol.
  2004, no.~15, pp. 2385--2395, 2004.

\bibitem{gtzan1}
U.~Marchand and G.~Peeters, ``{Swing ratio estimation},'' in \emph{Proc.
  Digital Audio Effects}, Trondheim, Norway, 2015, pp. 423--428.

\bibitem{gtzan2}
G.~{Tzanetakis} and P.~{Cook}, ``Musical genre classification of audio
  signals,'' \emph{IEEE Trans. Speech and Audio Processing}, vol.~10, no.~5,
  pp. 293--302, 2002.

\bibitem{carnatic}
A.~{Srinivasamurthy} and X.~{Serra}, ``A supervised approach to hierarchical
  metrical cycle tracking from audio music recordings,'' in \emph{Proc. IEEE
  Int. Conf. Acoust. Speech Signal Process.}, 2014, pp. 5217--5221.

\bibitem{beatles}
M.~E.~P. Davies, N.~D. Quintela, and M.~Plumbley, ``Evaluation methods for
  musical audio beat tracking algorithms,'' in \emph{Queen Mary University of
  London, Centre for Digital Music, Tech. Rep. C4DM-TR-09-06}, 2009.

\bibitem{robbiewilliams}
B.~D. {Giorgi}, M.~{Zanoni}, A.~{Sarti}, and S.~{Tubaro}, ``Automatic chord
  recognition based on the probabilistic modeling of diatonic modal harmony,''
  in \emph{Proc. Int. Workshop on Multidimensional Systems}, 2013.

\bibitem{rockdataset}
T.~de~Clercq and D.~Temperley, ``A corpus analysis of rock harmony,''
  \emph{Popular Music}, vol.~30, no.~1, p. 47–70, 2011.

\bibitem{Hockman2012}
J.~A. Hockman, M.~E.~P. Davies, and I.~Fujinaga, ``{One in the jungle: Downbeat
  detection in hardcore, jungle, and drum and bass},'' \emph{Proc. Int. Soc.
  Music Inf. Retr. Conf.}, pp. 169--174, 2012.

\bibitem{NEURIPS2019_90fd4f88}
M.~Zhang, J.~Lucas, J.~Ba, and G.~E. Hinton, ``Lookahead optimizer: $k$ steps
  forward, 1 step back,'' in \emph{Proc. Advances in Neural Information
  Processing Systems}, 2019, pp. 9597--9608.

\bibitem{Raffel14mir_eval:a}
C.~Raffel, B.~Mcfee, E.~J. Humphrey, J.~Salamon, O.~Nieto, D.~Liang, and
  D.~P.~W. Ellis, ``mir\_eval: a transparent implementation of common mir
  metrics,'' in \emph{Proc. Int. Soc. Music Inf. Retr. Conf.}, 2014, pp.
  367--372.

\bibitem{Davies2019}
M.~E.~P. Davies and S.~B{\"{o}}ck, ``{Temporal convolutional networks for
  musical audio beat tracking},'' in \emph{Proc. Eur. Signal Process. Conf.},
  2019.

\bibitem{vaswani2017attention}
A.~Vaswani \emph{et~al.}, ``Attention is all you need,'' in \emph{Proc.
  Advances in Neural Information Processing Systems}, 2017, pp. 5998--6008.

\bibitem{chiu21eusipco}
C.-Y. {Chiu}, J.~{Ching}, W.-Y. {Hsiao}, Y.-H. {Chen}, A.~W.~Y. {Su}, and Y.-H.
  {Yang}, ``Source separation-based data augmentation for improved joint beat
  and downbeat tracking,'' in \emph{Proc. Eur. Signal Process. Conf.}, 2021.

\end{thebibliography}

\end{document}